\documentstyle[12pt,amsbsy]{article}
\begin{document}
\setlength{\unitlength}{1mm}
\newcommand{\te}{\theta}
\newcommand{\bee}{\begin{equation}}
\newcommand{\ene}{\end{equation}}
\newcommand{\tra}{\triangle\theta_}

{\hfill Preprint DSF-3/94} \vspace*{1cm} \\
\begin{center}
{\Large\bf Radiation Damping and Quantum Excitation}\end{center}
\begin{center}
{\Large\bf for Longitudinal Charged Particle Dynamics}
\end{center}
\begin{center}
{\Large\bf in the Thermal Wave Model}\end{center}
\vspace{1cm}

\begin{center}
{{\large R. Fedele$^{1,2}$, G. Miele$^{1,2}$, and L. Palumbo$^{3,4}$}}
\end{center}

\vspace{1.5cm}

\noindent
$^{1}${\it Dipartimento di Scienze Fisiche, Universit\`a di Napoli
"Federico II", Napoli, Italy}\\
$^{2}${\it INFN Sezione di Napoli, Napoli, Italy}\\
$^{3}${\it Dipartimento di Energetica, Universit\`a "La Sapienza", Roma,
Italy}\\
$^{4}${\it INFN Laboratori Nazionali di Frascati, Frascati, Italy}

\vspace{1cm}

\begin{abstract}
On the basis of the recently proposed {\it Thermal Wave Model (TWM)
for particle beams}, we give a description of the longitudinal charge
particle dynamics in circular accelerating machines by taking into
account both radiation damping and quantum excitation (stochastic
effect), in presence of a RF potential well. The longitudinal dynamics
is governed by a 1-D Schr\"{o}dinger-like equation for a complex wave
function whose squared modulus gives the longitudinal bunch density
profile. In this framework, the appropriate {\it r.m.s. emittance}
scaling law, due to the damping effect, is naturally recovered, and
the asymptotic equilibrium condition for the bunch length, due to the
competition between quantum excitation (QE) and radiation damping
(RD), is found. This result opens the possibility to apply the TWM,
already tested for protons, to electrons, for which QE and RD are very
important.
\end{abstract}

\bigskip


\bigskip\bigskip\bigskip

\begin{center}{\it published in Phys. Lett. A194 (1994) 113-118.}\end{center}

\newpage

In the study of charged particle beam dynamics for accelerators and plasma
physics, a number of nonlinear and collective effects are relevant \cite{uno}.
Due to the
electromagnetic interactions between the particles and their image
charges induced in the surroundings, these nonlinear effects also
acquire
collective nature \cite{uno}. This property is enhanced for very intense beams,
which are employed in very high luminosity colliders.
In addition, radiation damping and quantum electromagnetic fluctuations
(quantum excitation) are generally present in the beam longitudinal dynamics,
and, in particular, for electron circular accelerating machines
are not negligible \cite{ventiquattro}.

Recently, a {\it Thermal Wave Model} (TWM) for charged particle beam
dynamics has been formulated \cite{sei} and successfully applied
to a number of
linear and nonlinear problems in beam physics \cite{sette}-\cite{dodici}.
In this approach, the beam transverse (longitudinal) dynamics is formulated
in terms of a complex function, the so called {\it beam wave function}
(BWF), whose squared modulus is proportional to the bunch
density. This wave function satisfies a Schr\"odinger-like equation in
which Planck's constant is substituted with the transverse
(longitudinal) bunch emittance \cite{sei},\cite{undici}.\\
In particular, this model is capable of reproducing the main results of
the conventional theory about transverse beam optics
and dynamics (in linear and nonlinear devices) \cite{sei}, and
it represents
a new approach for estimating the luminosity in particle accelerators
\cite{otto}, \cite{nove},
as well
as to study the self-consistent beam-plasma interaction \cite{sette}.
As far as the longitudinal bunch
dynamics is concerned, in this scheme on can describe, in a simple way,
the synchrotron motion
when both self-interaction and the radio frequency (RF)
potential well are taken into
account. In particular, the right conditions for the coherent
instability  in circular machines have been recovered
\cite{dieci},\cite{undici} and
soliton-like solutions for the beam density have been discovered
\cite{dieci}-\cite{dodici}.

In this letter we improve the {\it thermal wave model for longitudinal bunch
dynamics} given in \cite{dieci}-\cite{dodici}.
By starting from the conventional
longitudinal single-particle dynamics in circular accelerators,
the problem under study is formulated in terms of an
appropriate wave model which describes the evolution
of the beam, when the RF potential well is taken into account together
with radiation damping and quantum excitation. We show that
the longitudinal beam dynamics
is still correctly governed by a
Schr\"odinger-like equation for the BWF.
The envelope description is straightforwardly obtained from the wave
solution and, correspondingly,
the results are compared with those that are given in the conventional theory.
In particular, an asymptotic time-limit for the bunch length
and the r.m.s. emittance scaling law are obtained.

\vspace{0.5cm}

Let us consider the motion of a single particle within a
stationary bunch, travelling with velocity $\beta c$ ($\beta \approx 1$)
in a circular accelerating machine of radius $R_{0}=cT_{0}/2\pi$
($T_{0}$ being the revolution period). It is well-known that
if both radiation damping and quantum excitation are taken into account,
defining $x$ as the longitudinal displacement of the particle
with respect to the synchronous one, and $s\equiv ct\geq cT_{0}$
($t$ being the time),
the particle dynamics is governed by the following set of equations
\cite{sedici}:
\begin{equation}
\frac{dx}{ds}~ = ~\eta {\cal P}~~~,
\label{11}
\end{equation}
\begin{equation}
\frac{d {\cal P}}{ds }~ = ~ - \frac{q \Delta V}{c T_{0}
E_{0}}~-~\gamma {\cal P}~-~\frac{1}{E_{0}}\frac{dR}{ds}~~~,
\label{12}
\end{equation}
where ${\cal P}\equiv \frac{\Delta E}{E_{0}}$ denotes the relative
longitudinal energy spread of the single particle, computed with
respect to the synchronous one ($\Delta E=0$), and
$U(x,s)\equiv (1/cT_{0}E_{0})\int_{0}^{x}q~\Delta V(y)~dy$ is the
effective potential energy that the particle sees after a turn in the
ring ($\Delta V$ being the corresponding total voltage variation seen
by the particle). Moreover, in Eqs. (\ref{11})-(\ref{12})
$E_{0}$ and $q$ stand for the synchronous particle energy and charge
respectively, and $\eta \equiv
(1-\beta^2)-\alpha_{c}$ is the phase slip factor ($\alpha_{c}$
is the momentum compaction)\cite{uno}.\\
Finally, in (\ref{12}), $\gamma$ represents
the damping coefficient \cite{sedici} and $dR/ds$
accounts for the quantum excitation effect (noise),
$R(s)$ being the difference between the energy effectively radiated by the
particle during the time interval $s/c$ and the average of this energy.
Since $R(s)$ is a stochastic quantity, namely its average value
vanishes whilst its r.m.s. is not zero, we cannot treat the force
term $dR/ds$ as the
other ones of r.h.s. in (\ref{12}), which on the contrary have
deterministic nature. Hereafter, without loss of generality, we restrict
our analysis to electron machines ($q=e$, $|\eta|\approx \alpha_{c}$) for which
the e.m. emission is a relevant phenomenon.

The r.m.s. of $R^2 (s)$, $\overline{R^2}$, obeys to the following equation
\cite{ventiquattro}
\begin{equation}
\frac{d\overline{R^2}}{ds}~+~\gamma \overline{R^2}~=~Q_{p}~~~,
\label{13}
\end{equation}
with
\begin{equation}
Q_{p}~\equiv~\frac{\rho_{0}}{R_{0}}\frac{N_{\gamma}}{c} <u^{2}>~=~
\frac{55}{24\sqrt{3}}\frac{\hbar cr_{e}E_{0}\gamma_{0}^6}{R_{0}\rho_{0}^2}~~~,
\label{130}
\end{equation}
where $\rho_{0}$ is the magnetic bending radius,
$<u^{2}>^{1/2}$ is the quantum fluctuations associated to the noise,
$N_{\gamma}$ is the mean rate of photon emission, $r_{e}$ is
the classical electron radius, $\hbar$ is the Planck's constant, and
$\gamma_{0}\equiv (1-\beta^2)^{-1/2}$ is the relativistic factor.

It is easy to see that (\ref{11}) and (\ref{12}) are the usual equations for
the longitudinal motion \cite{sedici} under the substitution $s=ct$ and ${\cal
P}=\Delta E/E_{0}$.

It is worth to point out that the distinction between deterministic and
stochastic force terms is a necessary requirement to properly construct the
wave equation ruling the collective behaviour of the beam. This difference
can be understood observing how the emittance, which also describes a
stochastic effect but related to the temperature of the system, is involved
in a wave equation (Schr\"odinger-like equation)\cite{sei}-\cite{dodici}.
In fact, despite of
deterministic terms, the emittance is the only quantity which plays the role
analogous to a diffraction parameter. In the framework of TWM this parameter is
involved in the quantization rules.

To understand how to include quantum excitation in the TWM description,
we first consider the simplified situation in which quantum excitation
is negligible. Then, we generalize our results by taking into account
also this effect. Under this hypothesis,
by considering a linearized RF-voltage only
($U(x,s)=U_{RF}(x)
\approx (K/2 \eta)~x^{2}$, where $K$ is the RF cavity strength, supposed
for simplicity to be positive), it is easy to prove that the Lagrangian
associated to (\ref{11}) and (\ref{12}) is given by \cite{diciasette},
\cite{diciotto}
\begin{equation}
{\cal L}(x,x',s)=\frac{1}{2\eta} \left[x'^{2}
- K~x^{2}\right]~e^{\gamma s}~~~,
\label{14}
\end{equation}
where $x'\equiv dx/ds$.
Hence, computing the $x$-variable conjugate momentum
by $p \equiv \partial {\cal L}/\partial x'=(x'/\eta)
\exp(\gamma s)$, the corresponding hamiltonian results to be
\begin{equation}
{\cal H}(x,p,s)=\frac{\eta}{2}~ p^{2}~ e^{-\gamma s}+
\frac{1}{2}\frac{K}{\eta}~ x^{2}~ e^{\gamma s}~~~.
\label{23}
\end{equation}
In order to write a Schr\"odinger-like equation for the BWF, which
describes the longitudinal dynamics of a short bunch ($\sigma<<R_{0}$)
in presence of the only radiation damping we
have to follow the {\it Thermal Quantization Rules} (TQM),
in complete analogy with our previous works \cite{sei}-\cite{dodici} :
$p~\rightarrow\widehat{p}\equiv -i\epsilon
\partial/\partial x$~, and~~${\cal H}\rightarrow {\widehat{\cal H}}
\equiv i\epsilon \partial/\partial s$.
Therefore, (\ref{23}) gives (for $\eta\neq 0$)
\begin{equation}
i\eta \epsilon~ e^{-\gamma s} \frac{\partial}{\partial s}
\Psi(x,s)
{}~=~-\frac{\eta^2\epsilon^2~e^{-2 \gamma s}}{2}~\frac{\partial^2 }{\partial
x^2}\Psi(x,s)
{}~+~\frac{1}{2}K x^2 ~\Psi(x,s)~~~,
\label{25}
\end{equation}
where $\epsilon$ is a constant to be determined, which, according to
Ref.s \cite{sei}-\cite{dodici}, accounts for thermal spreading of the
bunch and plays the role fully similar to Planck's constant.
The BWF $\Psi(x,s)$ satisfies the following normalization condition
\begin{equation}
\int_{-\infty}^{\infty}|\Psi(x,s)|^{2}~dx~=~1~~~,
\label{25a}
\end{equation}
which, fixed for $s=0$,
holds for any $s$, due to hermiticity
of the hamiltonian operator (\ref{23}). Thus,
according to the features of the TWM, if $N$
is the total number of particles in the bunch, $N|\Psi (x,s)|^2$
represents the longitudinal bunch number density (number of particles
per unity length).\\
Interestingly, a complete set of solutions of (\ref{25}) is given in terms of
Hermite-Gauss modes:
\begin{eqnarray}
\Psi_{m}(x,s)  =
\frac{1}{\left\{2\pi2^{2m}(m!)^{2} \sigma^2(s)\right\}^{1/4}}~
\exp\left\{-\frac{x^2}{4 \sigma^2(s)}\right\}H_{m}\left(\frac{x}
{\sqrt{2}\sigma(s)}\right)
\nonumber\\
\times \exp\left\{i \frac{x^2}
{2 \eta\epsilon ~\exp(-\gamma s)~\rho(s)}+
i (1+2m) \phi(s)
\right\}~~~,
\label{30e}
\end{eqnarray}
where the $H_{m}$'s are the Hermite polynomials ($m=0,1,2,...$), and the
functions $\sigma(s)$, $\rho(s)$ and $\phi(s)$
satisfy the following set of differential equations
\begin{equation}
\frac{d^2 \sigma(s)}{d s^2}+\gamma \frac{d~\sigma(s)}{d s}+K\sigma (s)
-\frac{\eta^2\epsilon^2~e^{-2\gamma s}} {4~\sigma^{3}(s)}=0~~~.
\label{41a}
\end{equation}
\begin{equation}
\frac{1}{\rho}=\frac{1}{\sigma(s)}
\frac{d \sigma(s)}{d s}~~~,
\label{30a}
\end{equation}
\begin{equation}
\frac{d\phi}{d s}=-\frac{\eta\epsilon~e^{- \gamma s} }
{4 \sigma^{2}(s)}~~~~.
\label{30b}
\end{equation}
In particular, taking (\ref{30e}) for $m=0$
(fundamental mode), we note that $\sigma(s)$ results to be the
corresponding bunch length, defined as
\begin{equation}
 \sigma^2(s) = \int^{+\infty}_{-\infty} x^2~~ |\Psi_{0}(x,s)|^2~dx
\equiv <x^2> ~~~.
\label{30ca}
\end{equation}
In addition it is useful to define the momentum spread $\sigma_{p}(s)$
\begin{equation}
\sigma_{p}^2(s) = \epsilon^2\int^{+\infty}_{-\infty}
\left|\frac{\partial}{\partial x}\Psi_{0}(x,s)\right|^2~dx
\equiv <\widehat{p}^2> ~~~.
\label{30cab}
\end{equation}

By using the definition of $\widehat{p}$ and of
$\Psi_{0}(x,s)$, and the quantum formalism for the average of
the operators,
we can show that the following expression $A(s)$ is a constant of
motion
\begin{equation}
A(s)\equiv
2\left[ <x^2> <\widehat{p}^2> - \left( { 1\over 2}
< x \widehat{p} +
\widehat{p} x >\right) ^2 \right]^{1/2} = \epsilon~~~,
\label{31}
\end{equation}
and it coincides with the diffraction parameter $\epsilon$ of the TWM.
Observe that (\ref{31}) is formally identical to the well-known
Robertson-Schr\"odinger uncertainty relation
\cite{diciannove},\cite{venti}, and
$\epsilon$, which is straightforwardly obtained taking the
minimum of this relation,
is the natural extension to the quantum-like description of one of the
Courant-Snyder invariants, well-known in particle
accelerators \cite{ventuno}(Poincar\'e-Cartan invariants, in classical
mechanics \cite{ventidue}).

{}From (\ref{31}), we can also derive
the scaling law for the following effective emittance
$\overline{\epsilon}(s)$ ({\it quantum-like r.m.s. emittance}),
defined in analogy to the classical definition of r.m.s. emittance
given by Lapostolle \cite{uno},\cite{ventitre}
\begin{equation}
\overline{\epsilon}(s) \equiv {2 \over |\eta|} \left[ <x^2> <x'^2>
- \left( { 1 \over 2} <  x x' +
x' x >\right)^2 \right]^{1/2} = \epsilon~e^{- \gamma s}~~~.
\label{32}
\end{equation}
The same scaling law can be extrapolated from Eq. (\ref{41a}) by comparing
it with the corresponding expression for the undamped case \cite{undici},
consequently the diffraction parameter of TWM
$\epsilon$ represents the initial value $\overline{\epsilon}(0)$.
Furthermore, by virtue of (\ref{32}), the envelope equation
(\ref{41a}) could be rewritten substituting $\epsilon\exp(-\gamma s)$
with the effective emittance
$\overline{\epsilon}(s)$. It is worth to point out that the presence
of the friction-like term in (\ref{41a}) ($\gamma \neq 0$) is responsible
for the $s$-dependence (time dependence) of r.m.s. emittance, whereas
in the undamped case ($\gamma=0$) $\overline{\epsilon}$ is a constant of
motion and gives the accessible phase-space area associated with the system.

Remarkably, in order to obtain all the above results concerning with
radiation damping we could start from the following Schr\"odinger-like
equation
\begin{equation}
i\eta \overline{\epsilon}(s)~\frac{\partial}{\partial s}
\Psi(x,s)
{}~=~-\frac{\eta^2 \overline{\epsilon}^2(s)}{2}\frac{\partial^2 }{\partial
x^2}\Psi(x,s)
{}~+~\frac{1}{2}K x^2 ~\Psi(x,s)~~~,
\label{33}
\end{equation}
formally obtained by introducing in (\ref{23}) the following TQR:
$p\rightarrow\widehat{p}\equiv
-i\overline{\epsilon}(s)\partial/\partial x$~,
and~~${\cal H}\rightarrow {\widehat{\cal H}}
\equiv i\overline{\epsilon}(s) \partial/\partial s$,
where $\overline{\epsilon}(s)$ is given by (\ref{32}).
For the present case we note that:

a) $\overline{\epsilon}(s)$ satisfies the following differential equation
\begin{equation}
\frac{d\overline{\epsilon}}{ds}~+~\gamma\overline{\epsilon}=0
\label{330}
\end{equation}
with the initial condition $\overline{\epsilon}(0)=\epsilon$;

b) for this new TQR the Robertson-Schr\"odinger-like relation
(\ref{31}) does not give a constant of motion anymore, since it
results to be: $A(s)=\epsilon~e^{-\gamma
s}=\overline{\epsilon}(s)$.\\

The above remarks allow us to
immediately generalize our results to the case in which both
{\it deterministic} (RF potential well plus radiation damping)
and arbitrary {\it stochastic} effects are taken into account.
In this case, in fact, the system dynamics is assumed to be ruled by
\begin{equation}
i\eta \widetilde{\epsilon} (s)~\frac{\partial}{\partial s}
\Psi(x,s)
{}~=~-\frac{\eta^2 \widetilde{\epsilon}^2(s)}{2}\frac{\partial^2 }{\partial
x^2}\Psi(x,s)
{}~+~\frac{1}{2}K x^2 ~\Psi(x,s)~~~,
\label{33a}
\end{equation}
where now $\widetilde{\epsilon} (s)$ is
an arbitrary function of $s$, to be specified in correspondence of
the particular stochastic effects considered, but satisfying the
initial condition $\widetilde{\epsilon}(0)=\epsilon$.
Also in this more general
case of damped harmonic oscillator, we are able to give a
complete set of normalized solutions of Eq. (\ref{33a})
\begin{eqnarray}
\Psi_{m}(x,s)  =
\frac{1}{\left\{2\pi2^{2m}(m!)^{2} \sigma^2(s)\right\}^{1/4}}~
\exp\left\{-\frac{x^2}{4 \sigma^2(s)}\right\}H_{m}\left(\frac{x}
{\sqrt{2}\sigma(s)}\right)
\nonumber\\
\times \exp\left\{i \frac{x^2}
{2 \eta \widetilde{ \epsilon} (s) ~\rho(s)}+
i (1+2m) \phi(s)
\right\}~~~,
\label{34}
\end{eqnarray}
where now the function $\sigma(s)$ satisfies the following equation
\begin{equation}
\frac{d^2 \sigma(s)}{d s^2}+ \Gamma(s)
\frac{d~\sigma(s)}{d s}+K\sigma (s)
-\frac{\eta^2\widetilde{\epsilon}^2 (s)} {4~\sigma^{3}(s)}=0~~~,
\label{35}
\end{equation}
with $\Gamma(s) \equiv -d \log[\widetilde{\epsilon} (s)]/ds$,
and $\rho(s)$ and $\phi(s)$ have the same definition of ({\ref{30a})
and (\ref{30b}) in terms of $\sigma(s)$. By defining $\tilde{\sigma}(s)$ as
\begin{equation}
\tilde{\sigma}(s) \equiv \sigma(s)~\exp\left\{{ 1 \over 2} \int_{0}^{s}
\Gamma(s')ds' \right\}=
\sigma(s) ~\sqrt{{\epsilon \over \widetilde{\epsilon} (s)}}~~~,
\label{36}
\end{equation}
Eq. (\ref{35}) becomes
\begin{equation}
\frac{d^2 \tilde{\sigma}(s)}{d s^2}+\overline{K}(s)\tilde{\sigma} (s)
-\frac{\eta^2 {\epsilon}^2} {4~\tilde{\sigma}^{3}(s)}=0~~~,
\label{37}
\end{equation}
where $\overline{K}(s)\equiv K - { \Gamma^2(s) \over 4}
- { 1 \over 2} { d \Gamma(s) \over ds}$.\\
Some physical consideration are in order

1) At the early time ($\gamma s<<1$), $\widetilde{\epsilon}\simeq
\epsilon$, and $\sigma (s)\simeq \tilde{\sigma} (s)$. We physically expect
that, during this time scale, the damping rate $\Gamma (s)$ is maximum
whereas the quantum excitation is negligible. In fact, at the beginning,
due to the small number of produced photons, the photon noise is
negligible. Consequently, $\Gamma (s)\simeq \gamma =const.$,
and, thus, $\overline{K}(s)\simeq K-(\gamma^{2}/4)=const.$ In
particular, for $K > (\gamma^{2}/4)$,
$\overline{K}$ is a positive constant, so that $\tilde{\sigma} (s)$ and
$\sigma (s)$ are limited functions.

2) For very large time ($\gamma s>>1$), since in the usual description
of the circular accelerating machines \cite{ventiquattro} a sort of
asymptotic equilibrium is reachable,
we can seek this equilibrium by assuming that $\Gamma (s)$ vanishes as $s$
increases its values. This physically means that, as $s$ grows, the
radiation emission produces more and more photons which, in turn,
increase the quantum excitation. Since the radiation damping provides a
decreasing of the r.m.s. emittance, a sort of competition between
this damping and quantum excitation is thus established in such a
way to reach an asymptotic limit for $\widetilde{\epsilon} (s)$, say
$\epsilon_{D}$. Consequently, from the asymptotic condition $\Gamma
(\infty) =0$, it follows that $\overline{K}=K>0$. Thus, we conclude that,
for $\gamma s>>1$, $\tilde{\sigma} (s)$ is required to be limited,
and from (\ref{36})
and (\ref{37}) it follows that asymptotic equilibrium solutions
$\tilde{\sigma} (\infty)$ and $\sigma (\infty)$ for $\tilde{\sigma} (s)$ and
$\sigma (s)$, respectively, exist and results
\begin{equation}
\tilde{\sigma} (\infty)=\sigma
(\infty)\sqrt{\frac{\epsilon}{\epsilon_{D}}}~~~.
\label{37c}
\end{equation}

Since the present description holds for an arbitrary
form of $\widetilde{\epsilon}(s)$,
on the basis of the above physical considerations we
have to assume its explicit behaviour or, equivalently, to assume the
form of $\Gamma (s)$, in such a way to recover the
results of conventional theory \cite{ventiquattro}. To this end, we observe
that (\ref{330}),
which is valid in the case of negligible quantum excitation
($\epsilon_{D}=0$), suggests to assume in the more general case
$\epsilon_{D}\neq 0$, the following evolution equation for
$\widetilde{\epsilon} (s)$
\begin{equation}
\frac{d\widetilde{\epsilon}}{ds}~+~\gamma\widetilde{\epsilon}=
\gamma\epsilon_{D}~~~,
\label{37d}
\end{equation}
with $\widetilde{\epsilon} (0)=\epsilon$.
Consequently, we have
\begin{equation}
\widetilde{\epsilon} (s)~=~\epsilon~e^{-\gamma s}~+~\epsilon_{D}
(1-e^{-\gamma s})~~~,
\label{38}
\end{equation}
and
\begin{equation}
\Gamma (s)~=~\frac{(\epsilon -\epsilon_{D})\gamma}{(\epsilon
-\epsilon_{D})+\epsilon_{D} e^{\gamma s}}~~~.
\label{39}
\end{equation}
We observe that for $\gamma s>>1$, (\ref{33a}) becomes
\begin{equation}
i\eta \epsilon_{D}~\frac{\partial}{\partial s}
\Psi(x,s)
{}~=~-\frac{\eta^2\epsilon_{D}^2~}{2}~\frac{\partial^2 }{\partial
x^2}\Psi(x,s)~+~\frac{1}{2}K x^2 ~\Psi(x,s)~~~,
\label{40}
\end{equation}
and the uncertainty relation at the minimum gives
\begin{equation}
\sigma (\infty)\sigma_{p}(\infty)~=~\frac{\epsilon_{D}}{2}~~~,
\label{41}
\end{equation}
where $\sigma_{p}(\infty)\equiv \lim_{s \rightarrow\infty}
<{\hat{p}}^{2}>$.
Note that $\sigma_{p}(\infty)$ represents the expectation value of the
longitudinal momentum spread of the bunch at the asymptotic
equilibrium. This means that its explicit determination does not
depend on the history which led the system to the equilibrium condition, but it
depends on the quantum fluctuations (photon noise) only.
Furthermore, $\sigma (\infty)$ can be explicitly determined by
imposing the asymptotic equilibrium in the (\ref{35}). Thus
\begin{equation}
\sigma^2 (\infty)~=~\frac{|\eta|\epsilon_{D}}
{2\sqrt{K}}~~~,
\label{43}
\end{equation}
which, by using (\ref{41}), becomes
\begin{equation}
\sigma (\infty)~=~\frac{|\eta|R_{0}}{\nu_{s}}~\sigma_{p}(\infty)~~~,
\label{44}
\end{equation}
where $\nu_{s}$ (synchrotron number) \cite{uno} stands for the ratio
between $\Omega_{s}\equiv c\sqrt{K}$ \cite{uno} (synchrotron frequency)
and the revolution frequency $\omega_{0}=\beta c/R_{0}\approx c/R_{0}$.
Consequently,
since from (\ref{130}) the equilibrium value $(\overline{R^2})_{eq}$
of $\overline{R^2}$ is $Q_{p}/\gamma$, and since it is proved that
the equilibrium energy spread is given by $\frac{1}{2}(\overline{R^2})_{eq}$
\cite{ventiquattro}, it is very easy to recognize that
\begin{equation}
\sigma_{p}^2(\infty)~=~\frac{(\overline{R^2})_{eq}}{2E_{0}^2}~=~\frac{55}
{48\sqrt{3}}\frac{\hbar cr_{e}\gamma_{0}^6}{\gamma E_{0}R_{0}\rho_{0}^2}~~~.
\label{42}
\end{equation}
Hence, by combining (\ref{42}) and
(\ref{44}) we finally obtain
\begin{equation}
\sigma
(\infty)~=~\frac{|\eta|R_{0}}{\nu_{s}}\left[\frac{(\overline{R^2})_{eq}}
{2E_{0}^2}\right]^{1/2}~~~.
\label{45}
\end{equation}
Eq. (\ref{44}) recovers the well-known proportionality between the
equilibrium bunch length $\sigma (\infty)$ and the corresponding
equilibrium momentum spread $\sigma_{p}(\infty)$, given in literature,
whilst (\ref{42}) and (\ref{45}) give explicitly these values in
terms of the quantum fluctuations. We remark that, as
$\sigma (s)$ and $\sigma_{p}(s)$ go to the
asymptotic equilibrium values, $\widetilde{\epsilon}$ goes to the minimum
value $\epsilon_{D}$ of
the r.m.s. emittance starting from the initial value $\epsilon$,
according to the scaling law (\ref{38}). We
point out that the reduction of $\widetilde{\epsilon}(s)$ is due to the
bunch cooling produced by the radiation damping, and, consequently,
$\epsilon_{D}$ represents the limit value of $\widetilde{\epsilon}(s)$
for which the bunch thermalization is completed. In addition, by
combining (\ref{41}), (\ref{42}) and (\ref{45}) we immediately get the
expression of $\epsilon_{D}$ in terms of electron Compton's wavelength
$\lambda_{e}\equiv \hbar /m_{e} c$ ($m_{e}$ being the electron rest mass)
\begin{equation}
 \epsilon_{D} = \left[ \frac{55}{24\sqrt{3}}
\frac{|\eta| r_{e}\gamma_{0}^5}{\nu_{s}\gamma \rho_{0}^2}
\right]~\lambda_{e}~\equiv ~ \chi_{D}~\lambda_{e}~~.
\label{46}
\end{equation}
In table 1 we report the values of the factor $\chi_{D}$ for some
electron-positron circular machines. Note that the large values found for
$\chi_{D}$ imply $\epsilon_{D}>>\lambda_{e}$. This suggests that
at the equilibrium the quantum-like behaviour of the system as a whole
still corresponds to the Liouville regime (thermal equilibrium) and,
according to (\ref{41}), would represent a sort of macroscopical coherence.
On the other hand, the quantum limit of such a coherence
is, of course, recovered for $\epsilon_{D}\approx \lambda_{e}$ (Heisenberg
regime).

\vspace{0.5cm}

In this letter we have presented an extension of the recently proposed
{\it thermal wave model for particle dynamics} \cite{sei} to the
longitudinal motion in circular accelerating machines when both RD
and QE are taken into account. In this framework,
the particle dynamics in the presence of a RF potential well is
governed by a 1-D Schr\"odinger-like equation for a complex wave
function, whose squared modulus gives the longitudinal bunch profile.
We have shown that the solutions for the BWF of this problem are given
in terms of the well-known Gauss-Hermite modes. In particular, the fundamental
mode (lowest-energy mode) gives a pure Gaussian space-distribution for the
particles, and the corresponding envelope equation gives an asymptotic
value for the bunch length, which is expressed in terms of the quantum
fluctuations (noise). Correspondingly, according to (\ref{38}),
as the beam is cooling,
due to RD, the emittance goes to the equilibrium value $\epsilon_{D}$ which
represents the equilibrium limit and plays the role of thermalization
value.

In conclusion, the above results allow us to apply the thermal wave
model, already successfully applied to the undamped longitudinal dynamics
(protons) \cite{undici}, to the synchrotron electron motion in a more
accurate way, since in this case both radiation damping and quantum
excitation are not negligible. This occurs, for example, for electrons
in circular accelerating machines.

\newpage

{\bf Table 1.}

\vspace{.5cm}

\begin{tabular}{|c|c|c|c|c|c|c|}
\hline
& & & & & & \\
machines & $\rho_{0}$ ($m$) & $\gamma_{0}$ & $|\eta|\approx
\alpha_{c}$ & $\nu_{s}$ &
$\gamma$  ($m^{-1}$) & $\chi_{D}$ \\
& & & & & & \\
\hline
& & & & & & \\
DA$\Phi$NE & $1.4$ & $1.02\cdot10^{3}$ & $0.52\cdot10^{-2}$ & $7.63\cdot
10^{-3}$ & $3.74\cdot10^{-7}$ & $3.8\cdot10^{6}$ \\
& & & & & & \\
EPA & $1.43$ & $1.2\cdot10^{3}$ & $3.3\cdot10^{-2}$ & $1.55\cdot
10^{-3}$ & $1.0\cdot10^{-7}$ & $9.7\cdot10^{8}$ \\
& & & & & & \\
SLAC & $165$ & $1.8\cdot10^{4}$ & $2.44\cdot10^{-3}$ & $5.22\cdot10^{-2}
$ & $3.6\cdot10^{-7}$ & $3.4\cdot10^{7}$ \\
B-FACTORY & & & & & & \\
& & & & & & \\
LEP & $3.1\cdot10^3$ & $1.1\cdot10^{5}$ & $2.9\cdot10^{-4}$ & $8.9\cdot
10^{-2}$ & $3.54\cdot10^{-7}$ & $5.7\cdot10^7$ \\
& & & & & & \\
\hline
\end{tabular}
\end{document}